\def\be{\begin{equation}}
	\def\ee{\end{equation}}
\def\ber{\begin{eqnarray}}
	\def\eer{\end{eqnarray}}
\def\bers{\begin{eqnarray*}}
	\def\eers{\end{eqnarray*}}

\documentclass[aps,prb,twocolumn,showpacs,groupedaddressamsmath,amssymb]{revtex4-2}
\usepackage{dcolumn}   
\usepackage{amsmath}
\usepackage{multirow}
\usepackage{graphicx}    
\usepackage{comment}
\usepackage{color}
\usepackage{makecell}
\usepackage{array}
\usepackage{ifthen}
\usepackage{float}
\usepackage[utf8]{inputenc}
\usepackage{hhline}
\usepackage{graphicx}
\usepackage{soul}
\usepackage[normalem]{ulem}
\usepackage{cancel}
\newboolean{includefigs}
\setboolean{includefigs}{true}      
\newboolean{includetext}
\setboolean{includetext}{true}     
\newcommand{\condcomment}[2]{\ifthenelse{#1}{#2}{}}

\begin{document}
	
	\title{Dimensional Confinement Driven Scattering Inversion in NaCrTe$_2$}
	\author{Himanshu Sharma$^1$, Bhawna Sahni$^2$, Tanusri Saha-Dasgupta$^3$ and Aftab Alam$^1$}
	\email{aftab@iitb.ac.in}
	\affiliation{$^1$Department of Physics, Indian Institute of Technology Bombay, Powai, Mumbai 400 076, India}
	\affiliation{$^2$School of Engineering, University of Warwick, Coventry, CV4 7AL, United Kingdom}
	\affiliation{$^{3}$Department of Condensed Matter and Materials Physics, S. N. Bose National Centre for Basic Sciences, JD Block, Sector III, Salt Lake, Kolkata, West Bengal 700106, India}
	\begin{abstract}
 Dimensionality reduction provides a powerful route to tune the electronic and magnetic properties of van der Waals materials, yet its influence on electronic transport remains complex due to competing effects from quantum confinement and modified scattering mechanisms. Here, we investigate this interplay in an antiferromagnetic semiconductor $\text{NaCrTe}_2$ using first-principles calculations combined with the Boltzmann transport equation beyond the constant relaxation time approximation. Our results show that the monolayer limit induces a coupled magnetostructural reconstruction, reducing the band gap from $0.44$ eV (bulk) to $0.15$ eV (monolayer) and significantly enhancing the static dielectric constant. This evolution triggers a fundamental scattering inversion: whereas bulk transport is limited by polar optical phonon (POP) scattering, the monolayer becomes dominated by acoustic deformation potential (ADP) scattering. We show that this crossover originates from the simultaneous suppression of the Fröhlich interaction through enhanced dielectric screening and the amplification of acoustic scattering due to pronounced lattice softening. These results clarify how the interplay between dielectric screening, lattice stiffness, and band topology governs transport in low-dimensional magnetic semiconductors, providing a framework to optimize their electronic performance.
	\end{abstract}
	\date{\today}
	
	\maketitle
	\section{Introduction}
Layered ternary chalcogenides of the $ABC_2$ family ($A$ = alkali metal, $B$ = transition metal, $C$ = chalcogen) have emerged as versatile platforms for exploring a wide range of electronic and magnetic phenomena, including thermoelectric transport, topological phases, and low-dimensional magnetism\cite{ACrTe2_Magnetic,Pan2013,Kurosaki2003}. Within this class, compounds such as $\text{AgBiSe}_2$ and $\text{TlBiTe}_2$ have been extensively investigated for their thermoelectric performance and topological properties, demonstrating how ternary chalcogenides can simultaneously host strong spin–orbit coupling and highly tunable band structures\cite{Pan2013,Kurosaki2003}. Previous studies on homologous ternary chalcogenides such as $\text{CuCrS}_2$ and $\text{AgCrSe}_2$ have revealed promising electronic and thermoelectric properties arising from their layered structures. In contrast, the transport physics of the corresponding alkali-metal counterparts remains largely unexplored, likely due to synthesis challenges\cite{Tewari2010,Damay2016,McKeever2023}. Recent investigations of alkali-based chalcogenide monolayers suggest that these systems can exhibit exceptionally high carrier mobilities\cite{Sun2020}. 
In their bulk form, van der Waals crystals such as $\text{NaCrTe}_2$ typically possess dispersive electronic bands within the strongly bonded layers, enabling efficient in-plane charge transport, while exhibiting relatively flat and heavy-mass bands along the stacking direction that limit the direction-averaged transport performance. Dimensionality reduction to the monolayer therefore removes the low-mobility out-of-plane transport channel and sharpens the density of states (DOS) near the band edges. 
However, such DOS enhancement alone does not necessarily guarantee improved transport properties. While localized states with larger effective mass can increase the Seebeck coefficient, a high DOS may simultaneously enhance carrier scattering, thereby reducing carrier mobility. Moreover, quantum confinement can modify both the band gap and the underlying bonding environment, further influencing the transport coefficients. For instance, in Cu-based systems such as CuSbS$_2$ and CuSbSe$_2$, interlayer decoupling in the monolayer leads to band-gap narrowing, which degrades the Seebeck coefficient and the overall thermoelectric power factor\cite{Cu_S_Cu_Se}.
    

Despite growing interest in ternary chalcogenides, a microscopic understanding of their transport physics remains limited. Most theoretical studies on related systems—from Cu-based compounds such as CuSbS$_2$, CuSbSe$_2$, $\text{CuAlTe}_2$, $\text{CuGaTe}_2$, and $\text{CuInTe}_2$ to Ag- and Au-based analogues including $\text{AuAlTe}_2$ and $\text{AgAlTe}_2$\cite{Cu_S_Cu_Se,Gudelli2015,Hasan2022,Kumar2013}—rely on the constant relaxation time approximation (CRTA), where the carrier scattering time is treated as an empirical constant (typically $\tau \approx 10$ fs). Although computationally convenient, CRTA neglects energy-dependent scattering processes and the role of scattering phase space, which can be strongly influenced by dimensionality. Similarly, studies on alkali-based counterparts such as $\text{LiCrTe}_2$\cite{LiCrTe2_Exp} mainly address magnetic anisotropy, leaving the connection between bonding asymmetry and carrier transport lifetimes largely unexplored.

Speaking about the low dimensional counterparts, some studies of monolayer systems of this family, including LiMTe$_2$ (M = Al, Ga, In) and bulk-to-2D comparisons in materials such as $\text{SnSe}$, $\text{MoS}_2$, and $\text{CdTe}$\cite{LiAl,Kaasbjerg2012,Li2023,Yadav2021} exists in the literature. These studies incorporate explicit scattering mechanisms such as deformation potential and Fröhlich interactions. However, they often attribute transport changes primarily to effective-mass renormalization or band-gap variations. Such treatments overlook the dual role of effective mass: while a large DOS effective mass favors a high Seebeck coefficient, a small conductivity effective mass is required to preserve high carrier mobility and competitive power factors. 
Moreover, dimensional reduction modifies the bonding symmetry, altering bond polarity and thereby affecting polar optical phonon (POP) interactions that dominate carrier scattering in the bulk. Confinement also reshapes the phonon dispersion\cite{SI}, influencing phonon group velocities and the acoustic deformation potential (ADP) scattering that becomes dominant in the monolayer limit. These considerations highlight the need for a detailed investigation that directly links confinement-induced symmetry breaking, the resulting redistribution of bond polarity, and the corresponding change in the dominant carrier scattering mechanisms.

Chromium-based tellurides of the form $A\text{CrTe}_2$ constitute another important subgroup of this family. In these materials, localized Cr $3d$ moments arranged on a triangular lattice give rise to layered antiferromagnetism, field-tunable magnetic phases, and pronounced magnetic anisotropy, as reported for compounds such as $\text{TlCrTe}_2$ and $\text{AgCrSe}_2$\cite{Baenitz2021,Ronneteg2005}. Recent studies further indicate that modifying the $A$-site cation or reducing dimensionality can significantly alter the balance between interlayer and intralayer exchange interactions, thereby affecting both the magnetic ground state and the underlying electronic dispersion\cite{Zhang2021_CrTe2_RT_FM,Abuawwad2023_CrTe2_TopoExchange,Yang2023_ACrTe2_dimensionality_comment}.
For example, experiments on $\text{LiCrTe}_2$ demonstrate that alkali intercalation can effectively tune the magnetic anisotropy and ordering temperature\cite{LiCrTe2_Exp}. More generally, replacing the $A$-site cation with a larger ionic species increases the interlayer spacing, which weakens the antiferromagnetic coupling between adjacent layers\cite{Hu2020_MnBi4Te7,Chen2024_IntercalationTuning}. These observations highlight the sensitivity of magnetic interactions in $A\text{CrTe}_2$ compounds to structural and chemical tuning, making them promising candidates for investigating dimensionality-driven magnetic phenomena in van der Waals systems.

In this context, $\text{NaCrTe}_2$ is particularly attractive. It crystallizes in the trigonal $P\bar{3}m1$ structure with well-defined van der Waals gaps and exhibits an A-type antiferromagnetic ground state in the bulk, while recent studies suggest that dimensional confinement and surface engineering can stabilize ferromagnetic configurations in reduced dimensions\cite{Wang2021,Zhang2024,Sun2020}. Despite these intriguing characteristics, a systematic understanding of how dimensional reduction influences the magnetic ground state and the associated electronic transport properties of $\text{NaCrTe}_2$ remains largely unexplored. Addressing this issue is therefore crucial for elucidating the interplay between reduced dimensionality, magnetic ordering, and charge transport in layered van der Waals magnetic semiconductors.

To rigorously assess the physics of strong correlation and influence of dimensional confinement, we combine a fully \textit{ab initio} scattering formalism with an accurate description of the electronic structure. To obtain a reliable ground-state electronic structure, we employ the  hybrid HSE06 exchange--correlation functionals\cite{HSE06}, capturing intermediate-range van der Waals interactions, and reducing self-interaction errors through partial exact exchange\cite{Chen2020,Heyd2003}. This improved treatment of exchange and correlation significantly modifies the predicted magnetic and electronic properties. In the bulk phase, HSE06 enhances the local Cr magnetic moment ($\mu(\text{Cr})$) from $3.18\,\mu_B$ (PBE) to $3.43\,\mu_B$ and opens a semiconducting band gap of $0.44$~eV, while strengthening the induced spin polarization on Te atoms. Upon dimensional reduction, the A-type antiferromagnetic bulk ground state ($\mu(\text{Cr})$=$\pm 3.43\,\mu_B$) evolves into a robust intralayer ferromagnetic monolayer ($\mu(\text{Cr})$$\approx 3.41\,\mu_B$), accompanied by pronounced magnetic disproportionation between the chalcogen layers due to broken surface symmetry. This confinement-driven redistribution of spin density renormalizes the electronic structure, narrowing the HSE06 band gap to $0.15$~eV in the monolayer. By integrating these electronic benchmarks with energy- and temperature-dependent transport calculations, we demonstrate that carrier transport in the monolayer is governed by a complex interplay of confinement-induced density-of-states sharpening, modified bond polarity, reduced phonon group velocities, and altered scattering phase space. These results establish $\text{NaCrTe}_2$ as a prototypical magnetic van der Waals semiconductor for elucidating how dimensional confinement and magnetic ordering cooperatively reshape carrier scattering and transport in two-dimensional systems.

\section{Computational Details}
First-principles calculations were performed within Density Functional Theory (DFT) using the Vienna Ab initio Simulation Package (VASP) \cite{VASP_1, VASP_2} with the Projector Augmented Wave (PAW) method \cite{PAW_Blochl, PAW_VASP}. The electronic ground state was evaluated using a hierarchy of exchange--correlation functionals. Structural optimizations were first carried out using the Generalized Gradient Approximation (GGA) of Perdew, Burke, and Ernzerhof (PBE)\cite{PBE_GGA}. \textcolor{black}{While the Perdew--Burke--Ernzerhof (PBE) functional generally provides reliable ground-state structures, it suffers from self-interaction errors in transition-metal chalcogenides because of the localized nature of the $d$ orbitals\cite{Bae2021, Wexler2020_XC_Chalcogenides, Peng2018_DelocalizationError}. A common correction is the inclusion of a Hubbard $U$ term, which often improves the description of correlated systems\cite{Macke2024}.}
To treat on-site Coulomb interactions in localized Cr $3d$ orbitals, a Hubbard correction (PBE+$U$) with $U_{\text{eff}} = 3.0$ eV was applied\cite{Dudarev1998, Liechtenstein1995}. The meta-GGA SCAN functional with rVV10 nonlocal correlation was also tested, both with and without the Hubbard correction (SCAN+$U$) \cite{SCAN, rVV10}. Final electronic band gaps and effective masses were obtained using the hybrid HSE06 functional. Van der Waals corrections (DFT-D3 for PBE/HSE) and Spin--Orbit Coupling (SOC) were included to accurately capture interlayer binding and relativistic effects\cite{DFT_D3}.
The plane-wave kinetic energy cutoff was set to 500 eV. Brillouin zone sampling employed a $\Gamma$-centered Monkhorst--Pack $k$-mesh of $12 \times 12 \times 3$ for bulk and $16 \times 16 \times 1$ for the monolayer. A vacuum spacing of 22 $\text{\AA}$ along the $c$-axis was used in monolayer calculations to remove spurious periodic interactions. Electronic self-consistency convergence was set to $10^{-6}$ eV.

Phonon dispersion and dielectric properties were calculated using Density Functional Perturbation Theory (DFPT) in VASP\cite{Baroni2001, Gajdos2006}. These calculations yielded static and high-frequency dielectric constants ($\varepsilon_0, \varepsilon_\infty$) and Born effective charges ($Z^*$) with a tighter convergence criterion of $10^{-8}$ eV. For two-dimensional dielectric properties, a slab-averaged rescaling formalism was employed \cite{Laturia2018}. The monolayer dielectric constants were derived from supercell values ($\epsilon_{\text{SC}}$) as

\begin{equation}
\label{eq:ep_l}
\epsilon_{2\text{D},\perp} = \left[ 1 + \frac{c}{t} \left(\frac{1}{\epsilon_{\text{SC},\perp}} - 1\right)\right]^{-1};\  \epsilon_{2\text{D},\parallel} =
1 + \frac{c}{t}(\epsilon_{\text{SC},\parallel}-1)
\end{equation}

\noindent where $c$ is the supercell lattice parameter along the stacking direction and $t$ is the effective thickness of the $\text{NaCrTe}_2$ monolayer slab.

Phonon frequencies were calculated using the Phonopy package\cite{Phonopy}. Chemical bonding analysis via Crystal Orbital Hamilton Population (COHP) was performed with the LOBSTER code\cite{LOBSTER}. Charge transfer was evaluated using Bader charge analysis\cite{Bader}, while Fermi surface topology was analyzed using IFermi\cite{IFermi}.
Electronic transport properties were calculated using the AMSET (Ab initio Model for Scattering and Transport) package, which solves the Boltzmann Transport Equation (BTE) within the Relaxation Time Approximation (RTA)\cite{AMSET}. Total scattering rates included contributions from acoustic deformation potential (ADP), polar optical phonon (POP), and ionised impurity (IMP) scattering, using parameters such as deformation potential, elastic constants, dielectric tensors, and phonon frequencies obtained from first-principles calculations. \textcolor{black}{All the transport calculations are done under rigid band approximation (RBA) assuming that doping only tunes the Fermi level without changing the topology of bands\cite{RBA1,RB2}.}

For a parabolic band, the ADP scattering rate depends on lattice stiffness and the deformation potential\cite{Deformation_Potential},

\begin{equation}
\label{eq:ADP}
\tau^{-1}_{\text{ADP}} \propto
\frac{m^* k_B T D^2}{\rho v_s^2}
\end{equation}

\noindent where $m^*$ is the density-of-states effective mass, $D$ the deformation potential described as shift in bands with lattice displacements, $\rho$ the mass density, and $v_s$ the longitudinal sound velocity.

The POP scattering rate, dominant in polar crystals, scales with the Fr\"ohlich coupling constant,

\begin{equation}
\label{eq:POP}
\tau^{-1}_{\text{POP}} \propto
\omega_{\text{po}}
\left(
\frac{1}{\varepsilon_{\infty}} -
\frac{1}{\varepsilon_{0}}
\right)
\frac{1}{|\mathbf{k}-\mathbf{k'}|}
\end{equation}

\noindent where $\omega_{\text{po}}$ is the polar optical phonon frequency and the difference between $\varepsilon_0$ and $\varepsilon_{\infty}$ determines the coupling strength.

Ionized impurity (IMP) scattering was treated using the Brooks--Herring formalism\cite{Brooks_Herring, IMP1, IMP2}:

\begin{equation}
\label{eq:IMP}
\tau^{-1}_{\text{IMP}} \propto
\frac{1}{\epsilon_0^2}
\left[
\frac{N_{\text{imp}}}{|\mathbf{k}-\mathbf{k'}|+\beta^2}
\right]
\end{equation}

\noindent where $\beta$ is the inverse screening length and $N_{\text{imp}}$ is the impurity concentration.

The total scattering rate was finally obtained using Matthiessen’s rule

\begin{equation}
\label{eq:matthiessen}
\tau^{-1}_{\text{total}} =
\tau^{-1}_{\text{ADP}} +
\tau^{-1}_{\text{POP}} +
\tau^{-1}_{\text{IMP}}
\end{equation}

	\section{Results and discussions}
	\subsection{Crystal Structure}
	\begin{figure}[!t]
		\centering
		\includegraphics[width=1.0\linewidth,height=0.5\textheight,keepaspectratio]{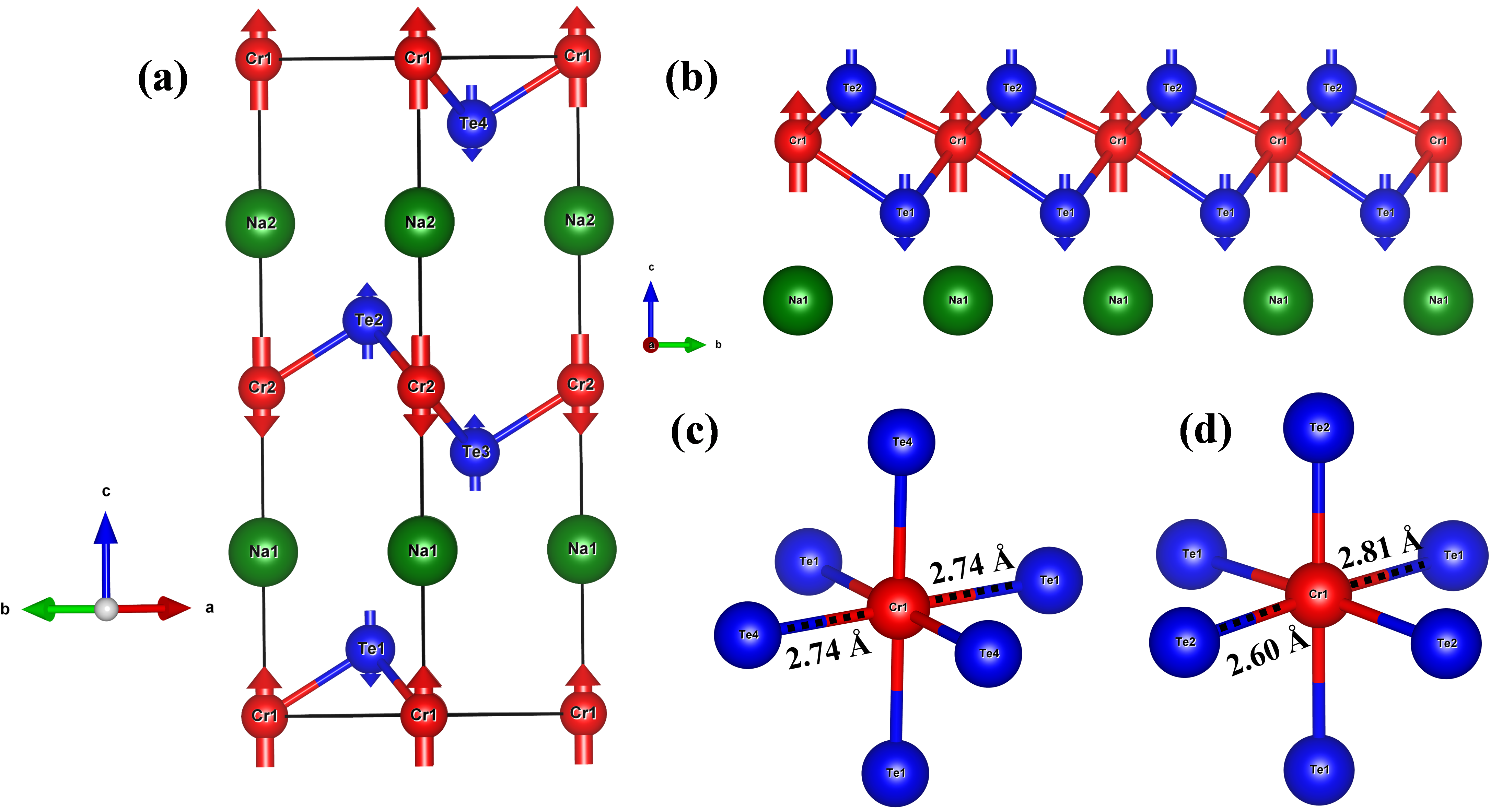}
		\caption{Crystal structure of (a) bulk  and (b) monolayer (side view) with edge shared octahedra of (c) bulk and (d) monolayer NaCrTe$_2$ respectively. }
		\label{fig:Crystal_structure}
	\end{figure}

The bulk phase of $\text{NaCrTe}_2$ crystallizes in a trigonal structure (space group $P\bar{3}m1$) composed of edge-sharing $\text{CrTe}_6$ octahedral layers separated by Na ions, as shown in Fig.~\ref{fig:Crystal_structure}(a). Structural optimization using the PBE functional yielded lattice parameters of $a = 4.075$ Å and $c = 14.646$ Å, which gives overestimated bond lengths ($>2.75$ Å). Including the DFT-D3 method with zero damping (capturing van der Walls (vdW) interaction) refined the lattice parameters to $a = 4.035$ Å and $c = 14.439$ Å, in good agreement with experimental value ($a \approx 4.005$ Å and $c \approx 14.88$ Å) \cite{Wang2021}. The PBE+vdW optimized structure was then used for subsequent magnetic and transport calculations. In the bulk phase, the $\text{CrTe}_6$ octahedra remain symmetric, with identical Cr--Te bond lengths of $2.748$ Å for both upper (Te$_1$) and lower (Te$_4$) chalcogen layers, reflecting preserved inversion symmetry.

To explore the two-dimensional counterpart, a monolayer was constructed based on the experimental feasibility of related compounds\cite{Zhang2024,Sun2020}. Previous studies show that the (001) surface of the isostructural $\text{NaCrS}_2$ can be mechanically exfoliated due to weak interlayer bonding\cite{Scheel1974}. Motivated by this, a $\text{NaCrTe}_2$ monolayer was modeled by isolating a single layer from the bulk and introducing a 22 Å vacuum along the $c$-axis to avoid spurious periodic interactions, as shown in Fig.~\ref{fig:Crystal_structure}(b). After relaxation ($P3m1$ symmetry), the monolayer preserves the in-plane lattice parameter $a = 4.035$ Å but undergoes significant internal reconstruction. The loss of inversion symmetry in the $\text{CrTe}_6$ unit results in distinct bond lengths: the Cr--Te$_2$ bond contracts to $2.60$ Å, while the opposite Cr--Te$_4$ bond elongates to $2.81$ Å. This structural distortion provides the geometric origin for modifications in the electronic and magnetic properties. In particular, the shorter Cr--Te$_2$ bond indicates enhanced hybridization on the surface-terminated side, which drives the magnetic moment and bond-strength disproportionation discussed in later sections. The dynamical stability of the reconstructed monolayer was verified through phonon dispersion calculations (see Fig.~SI\cite{SI} of SI\cite{SI}), which show no imaginary frequencies across the Brillouin zone, confirming its potential experimental synthesizability.


	\begin{figure}[!t]
		\centering
		\includegraphics[width=1.0\linewidth]{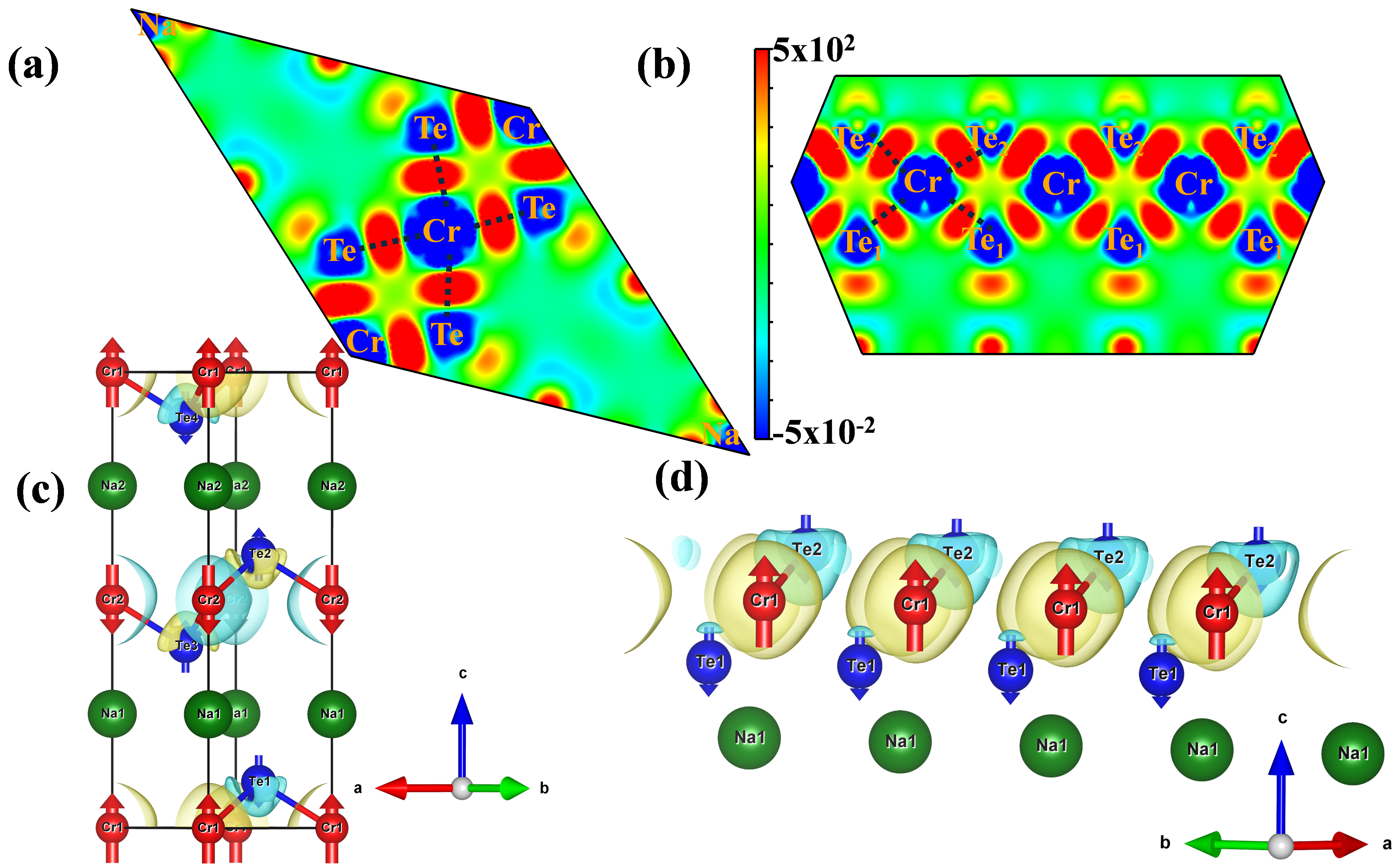}
		\caption{Charge density difference for (a) bulk and (b) monolayer $\text{NaCrTe}_2$, along with spin density distributions (isosurface = 0.005) for (c) bulk and (d) monolayer structures. In the charge-density plots, red regions indicate charge accumulation between atoms. In the spin-density maps, yellow and blue regions represent spin-up and spin-down polarization, respectively. }
		\label{fig:CG_SD}
	\end{figure}
	
	\subsection{Magnetic Properties}
NaCrTe$_2$ is experimentally reported to host an A-type antiferromagnetic (AFM) configuration in bulk, with ferromagnetically aligned Cr moments within each layer and antiferromagnetic coupling between neighbouring layers\cite{Zhang2024,Sun2020} (see Fig. \ref{fig:Crystal_structure}(a)). Our bulk calculations employ this A-type AFM ordering using a \(1\times1\times2\) supercell containing two Cr layers coupled antiferromagnetically. When a single layer is cleaved from the bulk, the interlayer superexchange pathway is removed, leaving only intralayer interactions. Consistent with previous theoretical studies on alkali–Cr–Te compounds, this stabilizes a ferromagnetic (FM) ground state in the monolayer \cite{Higo2016,Zhang2024} (see Fig. \ref{fig:Crystal_structure})(b). The calculated local magnetic moments for different exchange–correlation functionals are summarized in Table~\ref{tab:bulk_moment_full}, while results without SOC, together with additional band structures and densities of states, are provided in the SI (section II) \cite{SI}.
We examined several functionals—PBE, PBE+$U$, SCAN, SCAN+$U$, and HSE06—to assess the robustness of the magnetic state. Progressing from PBE to SCAN and HSE06 systematically increases the Cr moments (from \(3.18\,\mu_{\mathrm B}\) in PBE to \(3.27\,\mu_{\mathrm B}\) in SCAN and \(3.43\,\mu_{\mathrm B}\) in HSE06 for the bulk), reflecting reduced self-interaction errors and improved localization of the Cr $3d$ electrons. Applying a Hubbard \(U\) within PBE and SCAN further enhances the moments by localizing the $3d$ manifold. In the FM monolayer, the Cr moments remain comparable to the bulk values (about \(3.15\,\mu_{\mathrm B}\) for PBE and \(3.41\,\mu_{\mathrm B}\) for HSE06), indicating that dimensionality mainly alters interlayer coupling rather than the local Cr spin state.
	\begin{table*}[!htbp]
	\centering
	\caption{Local magnetic moments (in $\mu_{\mathrm{B}}$) on Cr and Te sites and band gap $E_g$ (in eV) for bulk $\text{NaCrTe}_2$ in the A-type AFM state.}
	\label{tab:bulk_moment_full}
	\setlength{\tabcolsep}{3.1pt}
	\resizebox{\textwidth}{!}{%
		\begin{tabular}{l | cc cc cc cc cc cc | cc}
			\hline\hline
			& \multicolumn{2}{c}{$\mu$(Cr1)} & \multicolumn{2}{c}{$\mu$(Cr2)} &
			\multicolumn{2}{c}{$\mu$(Te1)} & \multicolumn{2}{c}{$\mu$(Te2)} &
			\multicolumn{2}{c}{$\mu$(Te3)} & \multicolumn{2}{c}{$\mu$(Te4)} &
			\multicolumn{2}{c}{$E_g$} \\
			\hhline{~|------------|--} 
			Exch. Func. & w/o SOC & SOC & w/o SOC & SOC &
			w/o SOC & SOC & w/o SOC & SOC &
			w/o SOC & SOC & w/o SOC & SOC &
			w/o SOC & SOC \\
			\hline
			PBE    & 3.153 & 3.181 & $-3.153$ & $-3.181$ & $-0.115$ & $-0.124$ &  0.115 &  0.116 &  0.115 &  0.124 & $-0.115$ & $-0.116$ & 0.38 & 0.05 \\
			PBE+$U$  & 3.539 & 3.550 & $-3.539$ & $-3.550$ & $-0.197$ & $-0.201$ &  0.197 &  0.192 &  0.197 &  0.201 & $-0.197$ & $-0.192$ & 0.20 & metal \\
			SCAN   & 3.261 & 3.269 & $-3.261$ & $-3.269$ & $-0.157$ & $-0.160$ &  0.157 &  0.155 &  0.157 &  0.160 & $-0.157$ & $-0.155$ & 0.49 & 0.23 \\
			SCAN+$U$ & 3.608 & 3.623 & $-3.608$ & $-3.623$ & $-0.244$ & $-0.249$ &  0.244 &  0.243 &  0.244 &  0.249 & $-0.244$ & $-0.243$ & 0.28 & 0.015 \\
			HSE06  & 3.422 & 3.427 & $-3.422$ & $-3.427$ & $-0.191$ & $-0.195$ &  0.191 &  0.187 &  0.191 &  0.195 & $-0.191$ & $-0.187$ & 0.77 & 0.44 \\
			\hline\hline
		\end{tabular}%
	}
\end{table*}
  
A key qualitative change occurs on the Te ligands. In the centrosymmetric bulk, all Te atoms in the CrTe$_6$ octahedron are equivalent, leading to a symmetric charge distribution (Fig.~\ref{fig:CG_SD}(a)) and uniform induced moments arising from Cr–Te $p$–$d$ covalency, as seen in the spin-density isosurface of Fig.~\ref{fig:CG_SD}(c). Quantitatively, the induced moments on Te$_1$ and Te$_4$ are identical ($\approx -0.19\,\mu_B$ in HSE06). In the monolayer, however, structural confinement contracts the surface Cr–Te$_2$ bond to 2.60~\AA\ while elongating the inner Cr–Te$_1$ bond to 2.81~\AA. This distortion breaks charge symmetry (Fig.~\ref{fig:CG_SD}(b)), causing charge accumulation on the Te$_2$ site. The enhanced covalency on the shorter bond strengthens the magnetic exchange pathway, causing the surface Te$_2$ atom to develop a larger induced moment ($-0.27\,\mu_B$) compared to the bulk-like Te$_1$ site ($-0.14\,\mu_B$). This asymmetry is visible in the spin-density map shown in Fig.~\ref{fig:CG_SD}(d), where the spin lobes on Te$_2$ are more pronounced. Consequently, an out-of-plane asymmetric spin polarization emerges across the Te sublattice, which correlates with modifications of the band edges and transport channels discussed below.

	\begin{table*}[!htbp]
		\centering
		\caption{Local magnetic moments (in $\mu_{\mathrm{B}}$) on Cr and Te sites and band gap $E_g$ (in eV) for the ferromagnetic monolayer of $\text{NaCrTe}_2$, calculated with and without SOC for different exchange–correlation functionals. For the monolayer without SOC the spin-up and spin-down gaps are reported separately.}
		\label{tab:mono_moment_full}
		\setlength{\tabcolsep}{4pt}
		\resizebox{\textwidth}{!}{%
			\begin{tabular}{l | cc cc cc | ccc}
				\hline\hline
				& \multicolumn{2}{c}{Cr1} &
				\multicolumn{2}{c}{Te1} &
				\multicolumn{2}{c}{Te2} &
				\multicolumn{3}{c}{$E_g$} \\
				\hhline{~|------|---}
				Exch. Func. & w/o SOC & SOC &
				w/o SOC & SOC &
				w/o SOC & SOC &
				up (w/o SOC) & down (w/o SOC) & SOC \\
				\hline
				PBE    & 3.064 & 3.069 & $-0.075$ & $-0.075$ & $-0.150$ & $-0.146$ & 0.18 & 0.28 & 0.03 \\
				PBE+$U$  & 3.531 & 3.516 & $-0.156$ & $-0.156$ & $-0.280$ & $-0.264$ & 0.13 & 0.20 & metal \\
				SCAN   & 3.236 & 3.248 & $-0.121$ & $-0.126$ & $-0.231$ & $-0.225$ & 0.05 & 0.43 & 0.04 \\
				SCAN+$U$ & 3.637 & 3.653 & $-0.207$ & $-0.209$ & $-0.361$ & $-0.355$ & 0.007 & 0.46 & metal \\
				HSE06  & 3.409 & 3.413 & $-0.146$ & $-0.145$ & $-0.276$ & $-0.269$ & 0.37 & 0.54 & 0.15 \\
				\hline\hline
			\end{tabular}%
		}
	\end{table*}
	
	\subsection{Electronic structure}
	\subsubsection{Density of states}
\begin{figure}[b]
		\centering
		\includegraphics[width=1.0\linewidth]{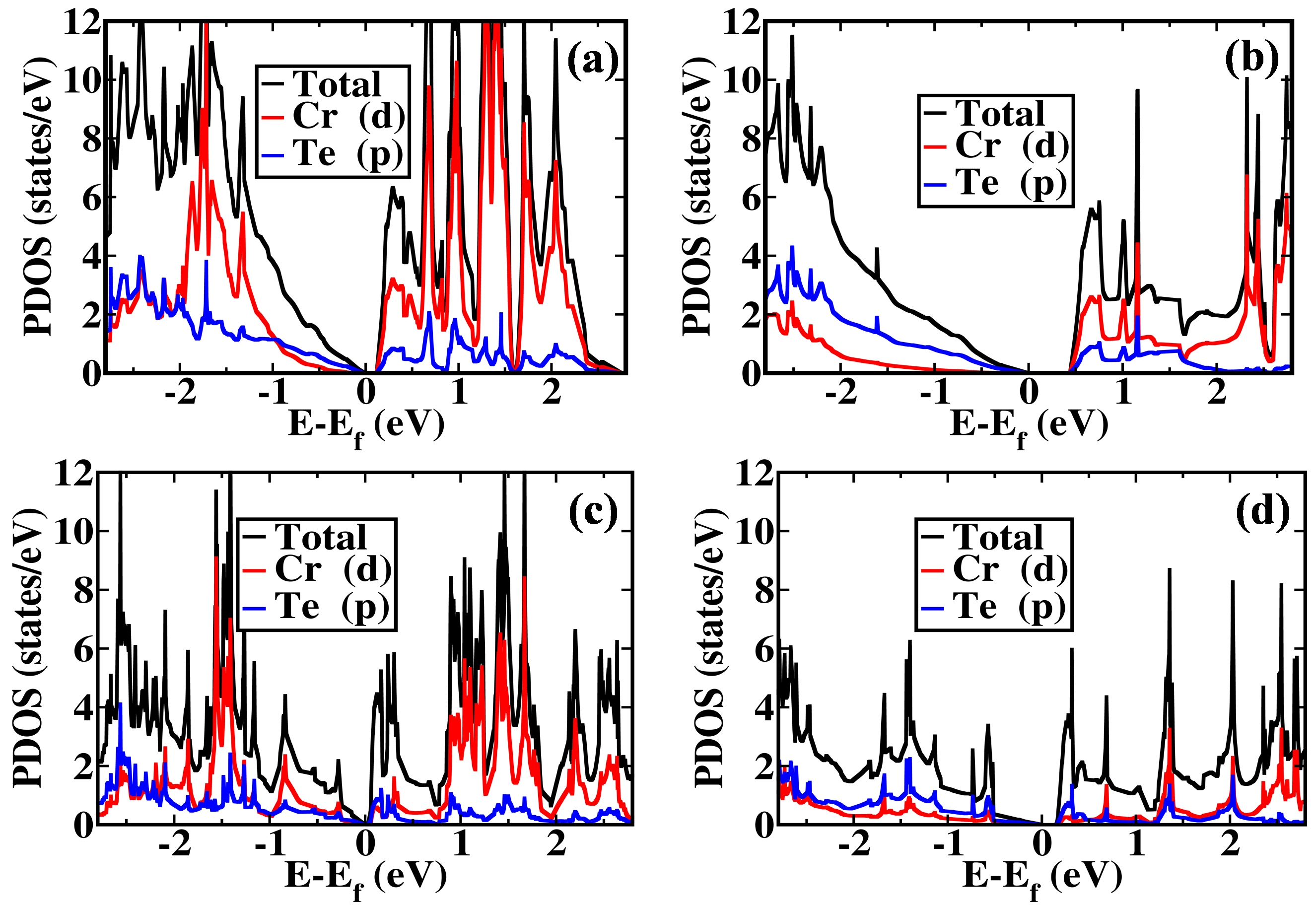}
		\caption{Atom projected density of states of bulk and monolayer NaCrTe$_2$ using (a,c) PBE (b,d) HSE06 functionals with SOC. }
		\label{fig:DOS}
	\end{figure}

The influence of the exchange–correlation functional on the orbital-resolved electronic structure of bulk $\text{NaCrTe}_2$ is summarized in the spin–orbit-coupled projected densities of states (PDOS) shown in Fig.~\ref{fig:DOS}(a) and \ref{fig:DOS}(b) for PBE and HSE06, respectively. PDOS obtained using PBE+$U$, SCAN, and SCAN+$U$ are provided in the SI\cite{SI}. Across all functionals, the system exhibits charge–transfer semiconductor character: the valence bands from about $-1$ eV up to the valence-band maximum are dominated by Te $5p$ states with a noticeable contribution from Cr $3d$, while the conduction bands are primarily Cr $3d$-like.
Within the PBE framework [Fig.~\ref{fig:DOS}(a)], strong hybridization between Cr $3d$ and Te $5p$ states occurs near the band edges, producing a narrow band gap. As discussed in the Supporting Information, introducing a Hubbard $U$ on the Cr $3d$ orbitals shifts the unoccupied $d$ states to higher energy and reduces the Cr contribution at the valence-band edge. This weakens $p$–$d$ hybridization and leads to a collapse of the gap, consistent with the nearly metallic behaviour observed in the PBE+$U$+SOC band structures (see section IV)\cite{SI}.

In contrast, the HSE06 hybrid functional [Fig.~\ref{fig:DOS}(b)] yields a qualitatively different electronic structure. Both Te $5p$ and Cr $3d$ features become sharper, the Te $p$ manifold shifts further below the Fermi level, and the onset of Cr $3d$ conduction states moves to higher energy compared to PBE, though remaining lower than in the +$U$ cases. This clearer separation between Te-dominated valence bands and Cr-dominated conduction bands produces the largest and well-defined bulk gap while preserving significant $p$–$d$ hybridization at the band edges. These results indicate that empirical +$U$ corrections tend to over-localize the Cr $3d$ states and artificially weaken Cr–Te covalency, whereas SCAN and especially HSE06 provide a more balanced description of localization and hybridization in chromium tellurides.

A similar trend is observed for the ferromagnetic monolayer, as shown in the PDOS of Fig.~\ref{fig:DOS}(c) and \ref{fig:DOS}(d), although confinement and symmetry breaking sharpen the spectral features. The electronic character remains charge–transfer-like, with Te $5p$ states forming the valence-band edge and Cr $3d$ states the conduction-band minimum. Compared with the bulk, reduced dimensionality results in narrower PDOS peaks and enhanced spectral intensity near the band edges. This is particularly evident in the HSE06+SOC result, where both the Te $5p$ valence and Cr $3d$ conduction manifolds appear as narrow, well-defined peaks at the gap edges, indicating an increased density of states relevant for transport.

The functional dependence in the monolayer closely mirrors that of the bulk. In the PBE+SOC case, Cr $3d$ and Te $5p$ states remain strongly hybridized near the Fermi level, producing a small semiconducting gap between a Te-dominated valence band and a mixed Cr–Te conduction band. Introducing a Hubbard $U$ [PBE+$U$+SOC] again shifts the unoccupied Cr $3d$ states upward and reduces overlap with Te $5p$ states, weakening $p$–$d$ hybridization and tending to close the gap. By contrast, SCAN and HSE06 recover stronger $p$–$d$ mixing at the band edges while maintaining a finite gap, consistent with their improved treatment of electronic localization.

\subsubsection{Bonding Analysis and Charge Distribution}

The bonding modifications induced by confinement are captured through charge-density difference and crystal orbital Hamilton population (COHP) analyses. The corresponding COHP plots obtained using PBE and HSE06 are provided in the SI. In the bulk, charge accumulation (red regions in Fig.~\ref{fig:CG_SD}(a)) is symmetrically distributed, consistent with uniform Cr–Te bond lengths and a moderate ICOHP of $-0.45$ eV. 
In the monolayer, however, contraction of the surface bond drives a pronounced electronic redistribution. As shown in Fig.~\ref{fig:CG_SD}(b), charge accumulation intensifies along the shorter Cr–Te$_2$ bond, indicating stronger orbital overlap. This observation is consistent with the calculated ICOHP values, which deepen to $-1.11$ eV for Cr–Te$_2$ compared to $-0.66$ eV for Cr–Te$_1$. The enhanced covalency also affects the ionic character: because electron density is concentrated in the bonding region, the effective charge transfer to the Te ligand is reduced. 

Bader charge analysis confirms this trend. In the bulk, equivalent Te atoms carry identical ionic charges of $+0.89e$, reflecting their uniform bonding environment. In contrast, the monolayer exhibits clear charge disproportionation, with the surface Te$_2$ site carrying only $+0.51e$ compared to $+1.10e$ on Te$_1$ (full data are provided in the SI \cite{SI}). This bonding asymmetry—particularly the reduced ionicity at the surface—forms the microscopic origin of the scattering mechanism inversion, as discussed later in the transport section.

	\begin{figure}[!b]
		\centering
		\includegraphics[width=1.0\linewidth]{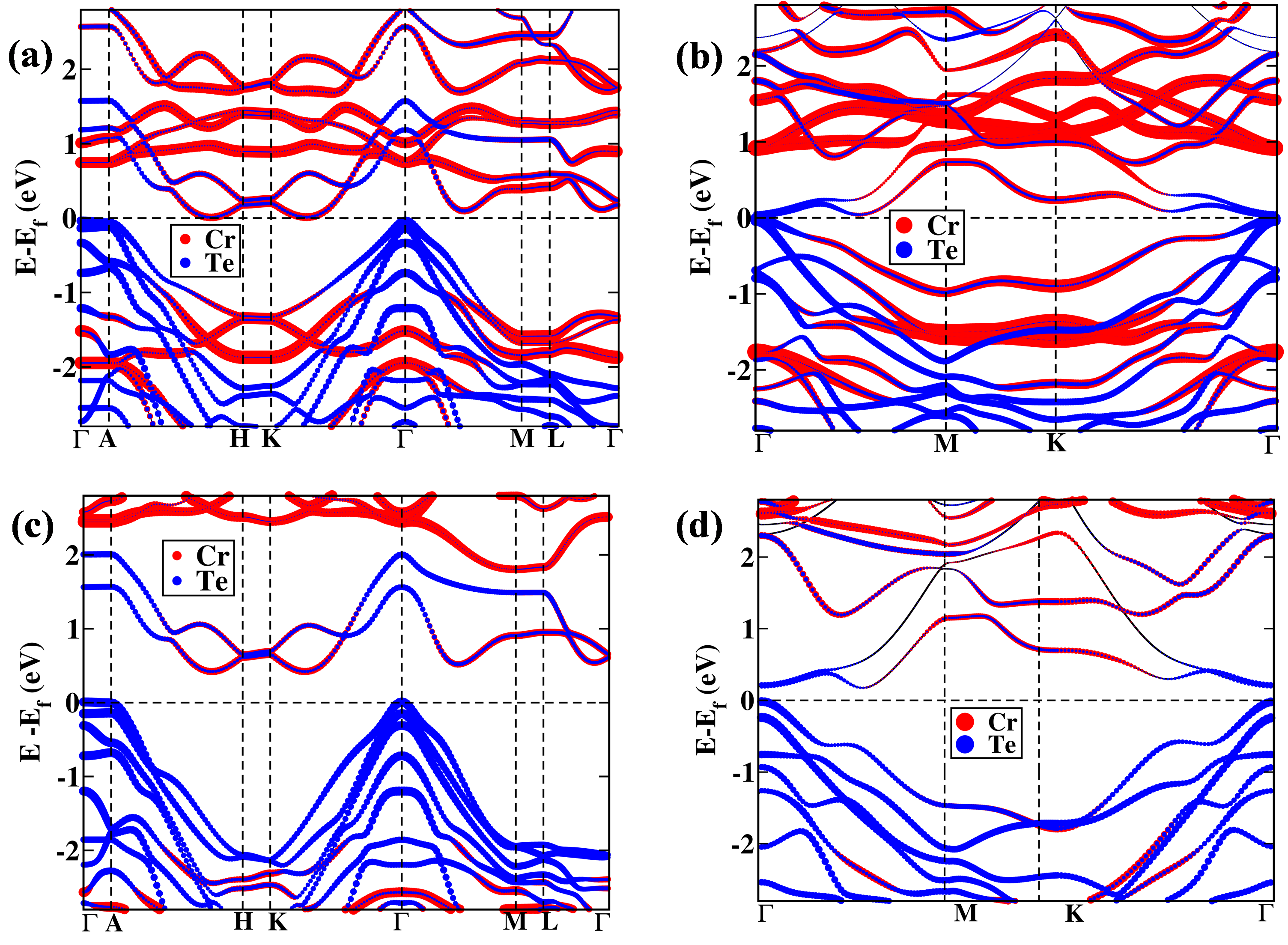}
		\caption{Atom projected band structure of bulk and monolayer NaCrTe$_2$ using (a,c) PBE (b,d) HSE06 functionals with SOC.  }
		\label{fig:Band_structure}
	\end{figure}

	\subsubsection{Band structure and Effective Masses}
The band dispersions for bulk and monolayer \(\text{NaCrTe}_2\) are shown in Fig.~\ref{fig:Band_structure}. Panels (a) and (c) display the bulk A-type AFM band structures calculated using PBE and HSE06 functionals including SOC, respectively, highlighting the Cr $3d$ and Te $5p$ characters. In the PBE+SOC case [Fig.~\ref{fig:Band_structure}(a)], \(\text{NaCrTe}_2\) exhibits a narrow indirect band gap of about 0.05~eV. The valence-band maximum (VBM) occurs at \(\Gamma\) and is dominated by Te $5p$ states, while the conduction-band minimum (CBM) lies along the in-plane \(\Gamma\)--M direction and is mainly Cr $3d$-like. Spin–orbit coupling lifts several degeneracies in the Te-derived valence manifold near \(\Gamma\) and along the \(\Gamma\)--K and \(\Gamma\)--M paths.

	\begin{figure*}[t]
		\centering
		\includegraphics[width=1.0\linewidth]{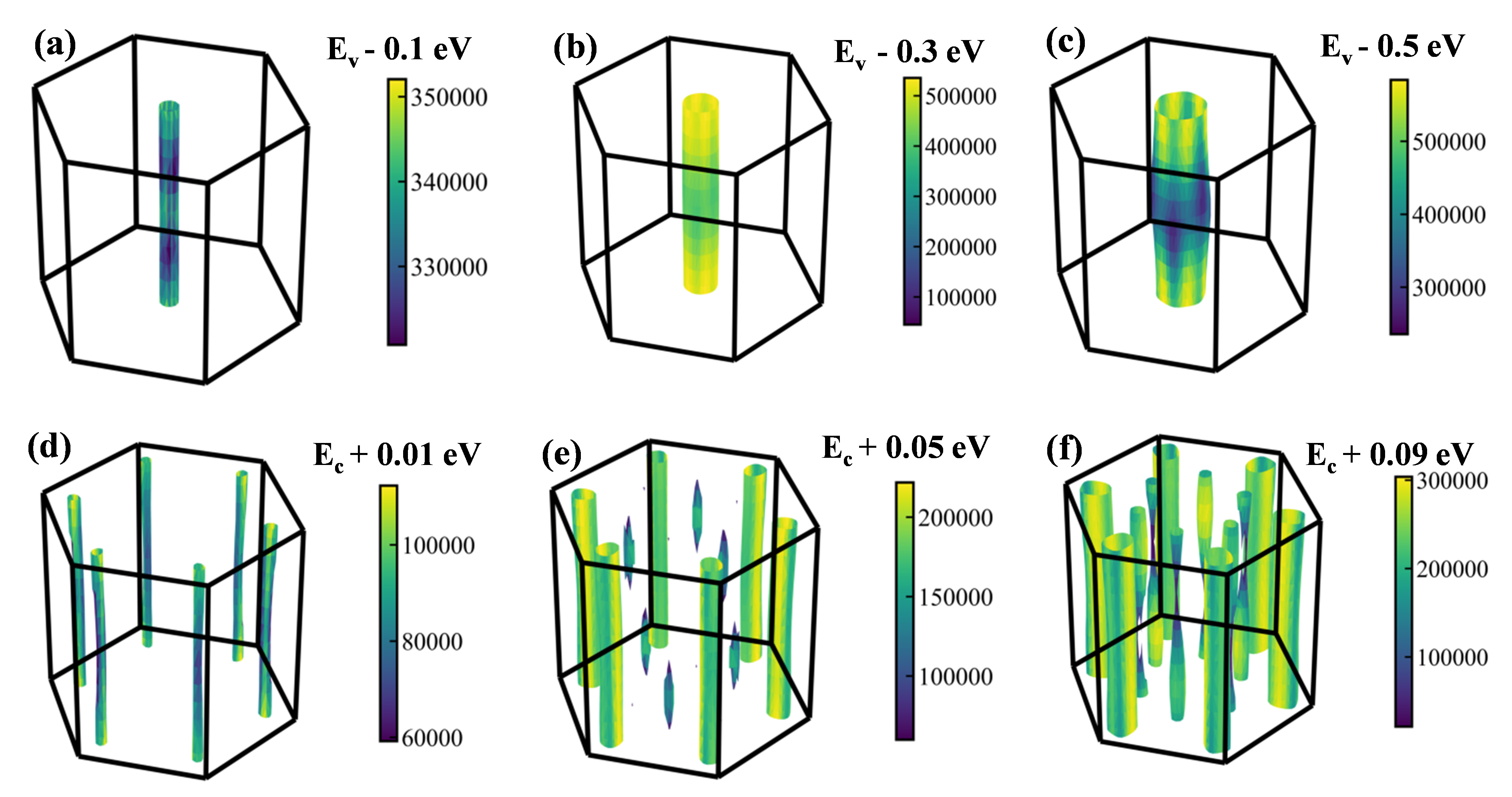}
		\caption{Isoenergy surface of the (a,b,c) valence band and (d,e,f) conduction band of bulk NaCrTe$_2$ at different chemical potentials. The colorbars signify the velocity.}
		\label{fig:Bulk_ISO}
	\end{figure*}

Within HSE06+SOC calculations [Fig.~\ref{fig:Band_structure}(c)], both the Te $5p$ valence bands and Cr $3d$ conduction bands shift away from the Fermi level, producing a larger indirect gap of 0.44~eV while preserving the VBM at \(\Gamma\) and the CBM along \(\Gamma\)--M. The conduction bands become less entangled and the Te- and Cr-dominated manifolds are more clearly separated, consistent with the sharper orbital-resolved features in the HSE06 PDOS [Fig.~\ref{fig:DOS}(b)]. Notably, the out-of-plane dispersion is much flatter than the in-plane dispersion, reflected in the heavy out-of-plane density-of-states effective mass (\(m_h^* \approx -3.14\,m_e\) out-of-plane versus \(-0.20\,m_e\) in-plane at the HSE06 level), highlighting the quasi-two-dimensional electronic character typical of layered chromium tellurides \cite{Vikram2022,Kutorasinski2015,Scheidemantel2011}.



Figures \ref{fig:Band_structure}(b,d) illustrate the electronic band structure of the ferromagnetic monolayer along the high-symmetry $\Gamma$–M–K–$\Gamma$ path. At the PBE+SOC level [Fig. \ref{fig:Band_structure}(b)], the monolayer behaves as a narrow-gap semiconductor with an indirect gap of approximately 0.03 eV. The valence-band maximum (VBM) at $\Gamma$ is dominated by Te $5p$ states, while the conduction-band minimum (CBM) between M and K is primarily Cr $3d$-like. This orbital character is consistent with the strong Cr–Te hybridization observed in the monolayer PDOS. Upon employing the HSE06+SOC functional [Fig. \ref{fig:Band_structure}(d)], the overall dispersion pattern is preserved, but the valence–conduction band separation increases to a direct gap of 0.15 eV. The observed bandgap reduction in the monolayer relative to the bulk can be attributed either to the selective renormalization of chalcogen states, as seen in $\text{CuSbS}_2$ and $\text{CuSbSe}_2$ \cite{Cu_S_Cu_Se} or to shifts in bond polarity similar to those observed in Li-disordered rock salt and doped 2D-$\text{SnS}_2$\cite{Li_rock, SnS2}. In our case, the bulk phase exhibits uniform and symmetric Cr–Te bond lengths, which corresponds to a lower degree of covalent character also supported by both the calculated ICOHP and charge transfer values (discussed in Sec. C2). This less covalent environment effectively stabilizes the wider bandgap of 0.44 eV. Conversely, dimensionality reduction to the monolayer triggers a structural reconstruction that significantly enhances the covalent nature of the Cr–Te bonds. This increased covalency facilitates more efficient charge transfer between the Te $5p$ and Cr $3d$ bands, thereby narrowing the charge-transfer energy ($\Delta$) and reducing the fundamental bandgap to 0.15 eV. Furthermore, confinement increases the density-of-states effective masses for both carriers ($m_e^* \approx 0.34\,m_e$ and $m_h^* \approx -0.28\,m_e$), which correlates with the enhanced density of states near the Fermi level. While this increased DOS may benefit thermopower, the heavier effective masses imply reduced intrinsic mobility, so the overall transport behaviour depends on the competition between confinement effects and carrier scattering, as discussed in the following section.

	\begin{figure*}[t]
		\centering
		\includegraphics[width=1.0\linewidth]{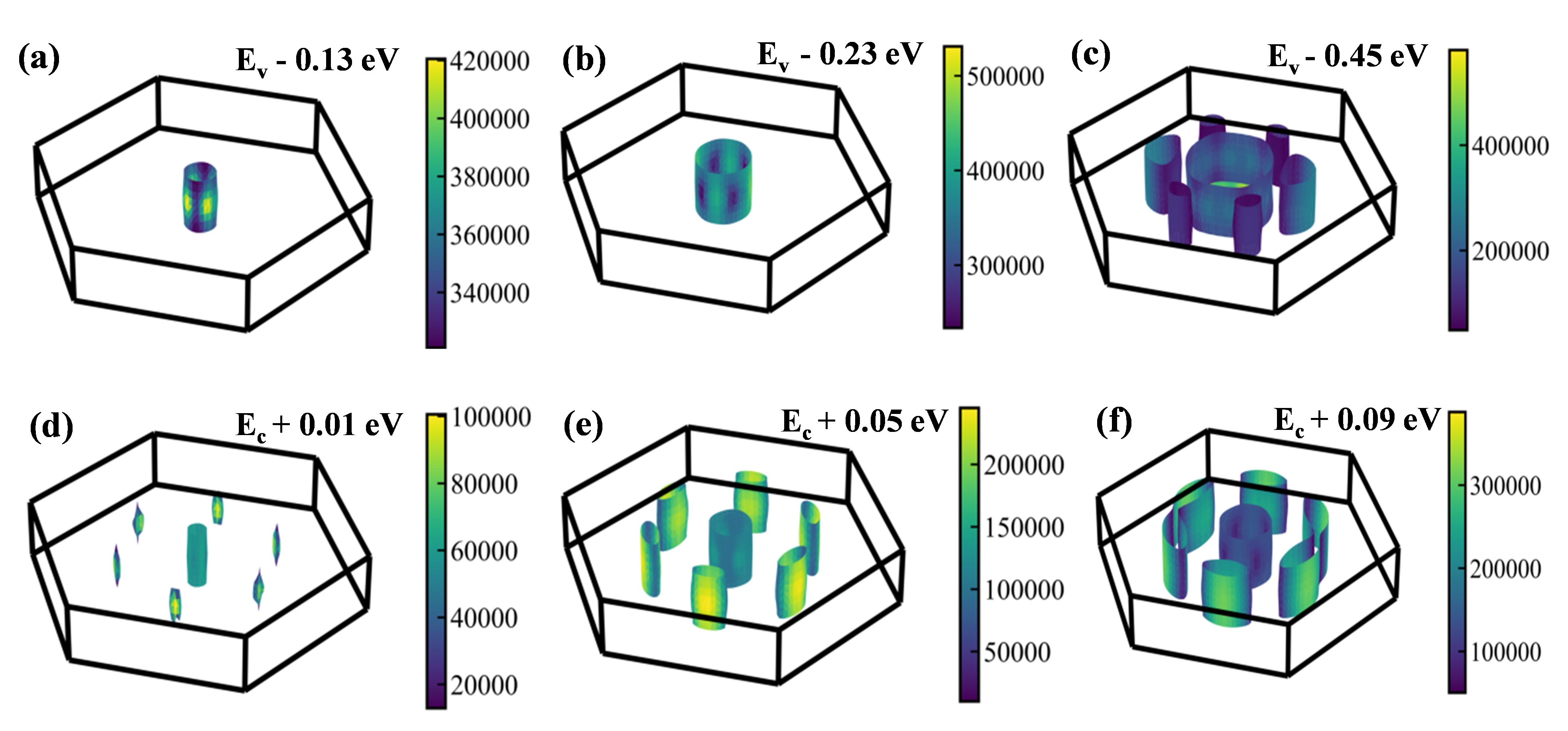}
		\caption{Isoenergy surface of the (a,b,c) valence band and (d,e,f) conduction band of monolayer NaCrTe$_2$ at different chemical potentials. The colorbars signify the velocity.}
		\label{fig:Mono_ISO}
	\end{figure*}
	
    \subsubsection{Fermi Isoenergy Surface}
 To elucidate the microscopic origin of the effective-mass renormalization, we plot the Fermi isoenergy surfaces of bulk and monolayer NaCrTe$_2$ in Figs.~\ref{fig:Bulk_ISO} and \ref{fig:Mono_ISO}, where color denotes the group velocity. In the bulk [Figs.~\ref{fig:Bulk_ISO}(a–c)], the valence-band isosurfaces form open cylindrical tubes along $k_z$, reflecting the layered electronic structure with strongly dispersive in-plane bands and weak out-of-plane dispersion. This topology results in large in-plane velocities ($\bar{v} \approx 3.3\times10^{5}$ m/s) and light hole masses. In contrast, the conduction-band isosurfaces [Figs.~\ref{fig:Bulk_ISO}(d–f)] consist of multiple pockets with lower velocities ($\bar{v} \approx 8.5\times10^{4}$ m/s) but a substantially larger total area (1.39 vs 0.41~\AA$^{-2}$ for holes), providing numerous parallel $n$-type transport channels.

In the monolayer [Fig.~\ref{fig:Mono_ISO}], removal of interlayer coupling only weakly modifies the electron isosurfaces [Figs.~\ref{fig:Mono_ISO}(d–f)], which evolve from compact pockets into dispersive sheets while retaining relatively low velocities ($\sim6\times10^{4}$ m/s), consistent with robust $n$-type transport. A pronounced change occurs in the valence band [Figs.~\ref{fig:Mono_ISO}(a–c)]: confinement suppresses $k_z$ dispersion and flattens the bands along M–K, transforming the cylindrical bulk surfaces into dense wall-like sheets. 

Remarkably, this dimensional reduction increases the average velocity by $\sim10\%$ (from $\sim3.3\times10^{5}$ to $\sim3.6\times10^{5}$ m/s) while reducing the isosurface area by nearly a factor of three (0.41 to 0.12~\AA$^{-2}$). This inverse correlation between velocity and isosurface area provides a clear signature of confinement-induced renormalization of the conductivity effective mass ($m^*_{\sigma}$). The same trend is reflected in the carrier mobility of the bulk and monolayer systems (see Sec.~VII of the SI~\cite{SI}). Although the large density of states enhances the scattering phase space for holes, their high velocity in the monolayer compensates for reduced relaxation times and ultimately contributes to the large electrical conductivity discussed in the next section.

\subsection{Transport properties}
\begin{figure}[t]
		\centering
		\includegraphics[width=1.00\linewidth]{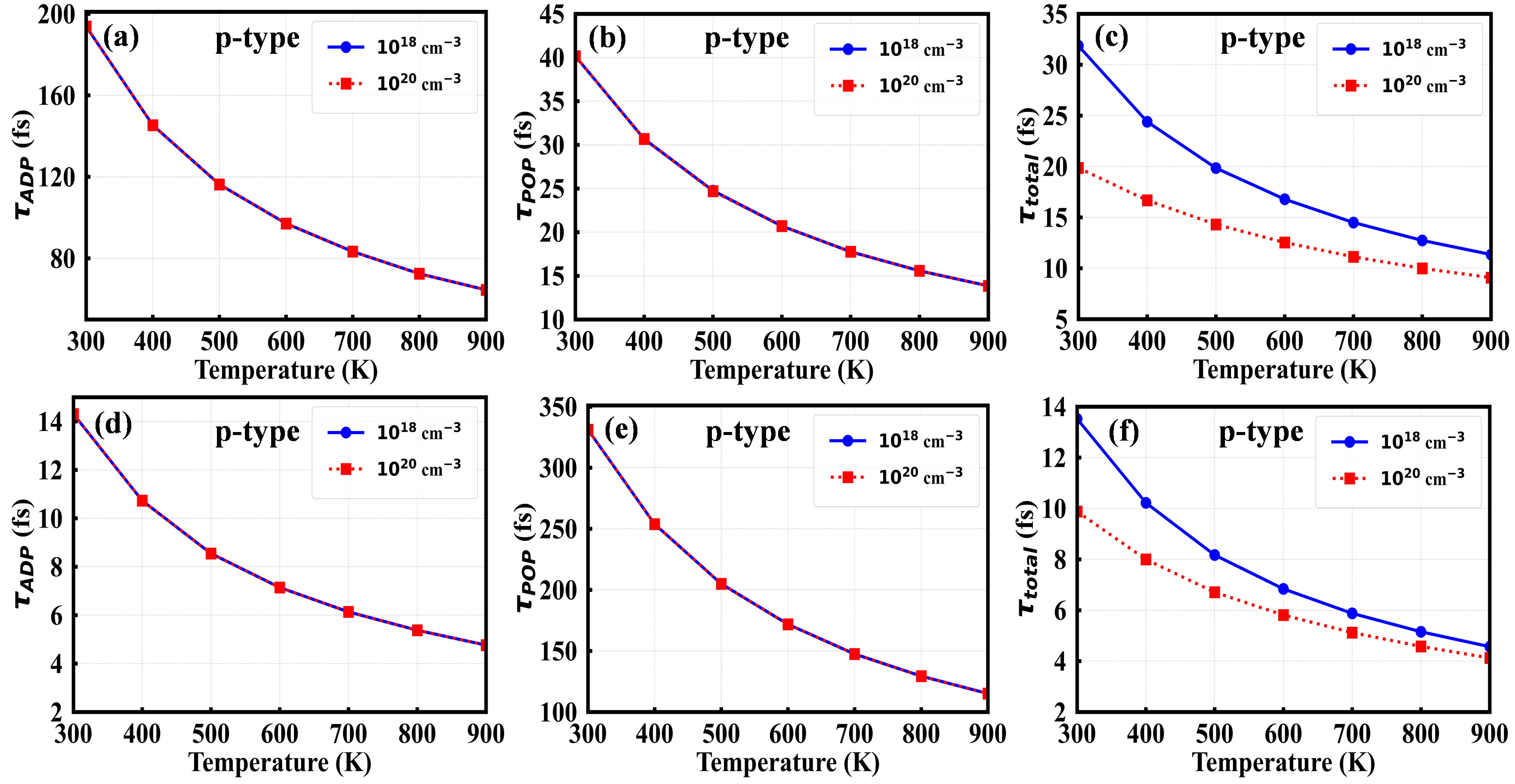}
		\caption{Relaxation time for different scattering mechanisms, ADP, POP and total, for (a,b,c) bulk and (d,e,f) Monolayer p-type NaCrTe$_2$. }
		\label{fig:Tau_p}
	\end{figure}

\begin{figure}[t]
		\centering
		\includegraphics[width=1.01\linewidth]{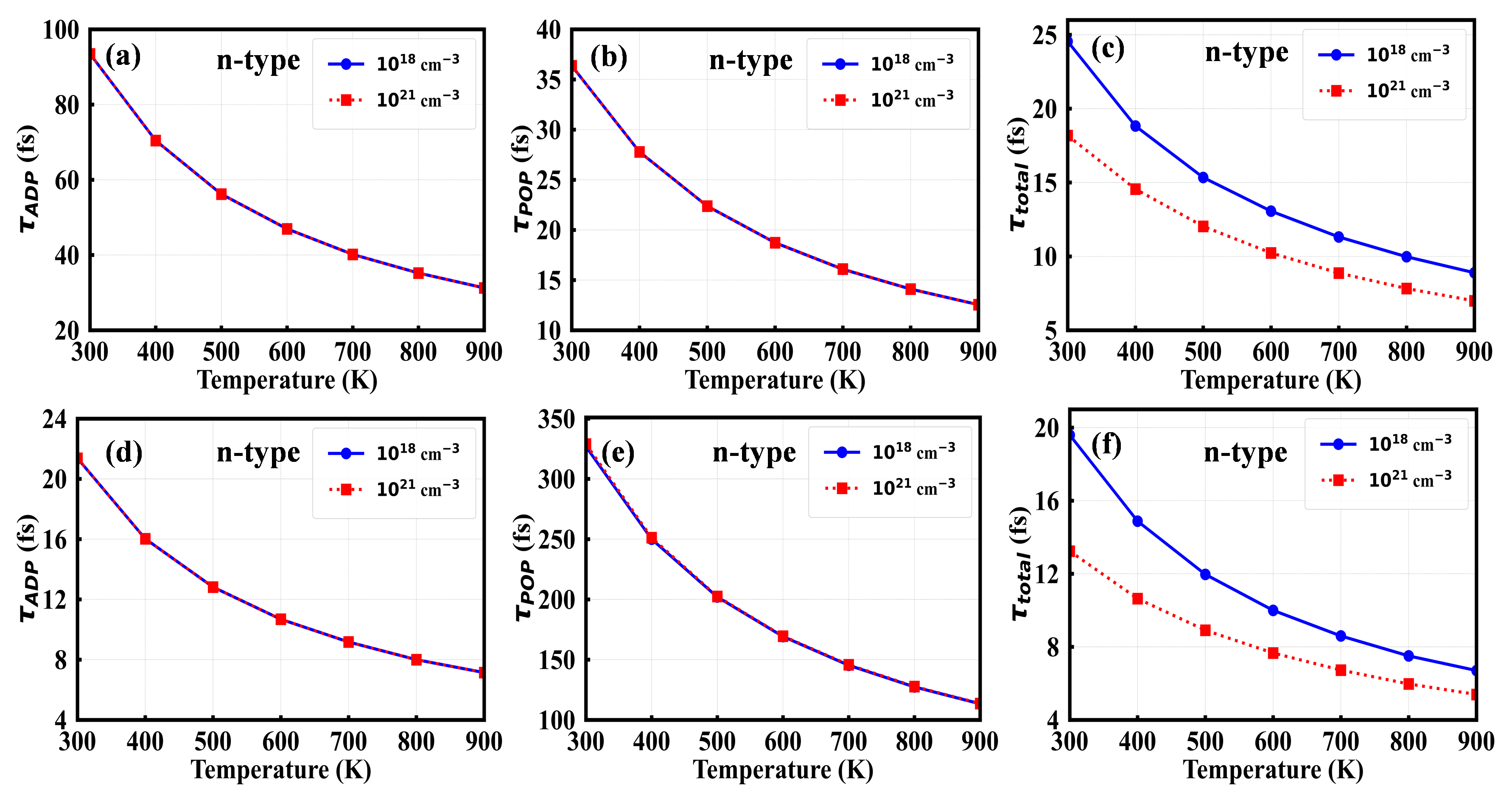}
		\caption{Relaxation time for different scattering mechanisms, ADP, POP and total, for (a,b,c) bulk and (d,e,f) Monolayer n-type NaCrTe$_2$.}
		\label{fig:Tau_n}
	\end{figure}

To identify the microscopic origin of the evolution of transport properties from bulk to monolayer, we analyze three key contributions to the carrier scattering rates: (i) the electron–phonon coupling strength described by the deformation potential, (ii) lattice stiffness represented by the phonon group velocity, and (iii) the long-range Fröhlich interaction characterized by the dielectric response $\left[\alpha \propto \left(\frac{1}{\varepsilon_\infty}-\frac{1}{\varepsilon_0}\right)\right]$ and Born effective charges. The definition and other details about these parameters are summarized in Sec.~V of the SI~\cite{SI}. 
The dimensional transition from bulk to monolayer introduces a major competition between the acoustic deformation potential (ADP) and polar optical phonon (POP) scattering channels. Figures~\ref{fig:Tau_p} and \ref{fig:Tau_n} present the temperature dependence of the relaxation times $\tau_{ADP}$, $\tau_{POP}$, and the total relaxation time $\tau_{total}$ for both low and high carrier concentrations regime. Because ADP and POP dominate the scattering processes and depend weakly on carrier concentration, ionized impurity scattering contributes only minor corrections to the total relaxation time (see Sec.~SVII of the SI~\cite{SI}).

In bulk NaCrTe$_2$, ADP and POP scattering dominate both $n$- and $p$-type transport, with holes exhibiting a larger deformation potential than electrons. When the system is reduced to a monolayer, structural reconstruction significantly modifies the lattice dynamics. The absence of interlayer coupling softens the acoustic phonons, reducing the longitudinal acoustic velocity from $\sim 3062$ m/s in the bulk to $\sim 2126$ m/s in the monolayer (Table~SI). Since the scattering rate scales inversely with the square of the sound velocity ($\tau^{-1}\propto v_s^{-2}$), this acoustic softening substantially enhances carrier scattering in the two-dimensional limit. 
Concurrently, chemical bonding becomes stronger in the monolayer. The integrated crystal orbital Hamilton population (ICOHP) for the Cr–Te bond becomes significantly more negative ($-1.11$ and $-0.66$ eV) compared to the bulk ($-0.45$ eV), indicating enhanced covalency. This stronger bonding increases the sensitivity of the band-edge electronic states to lattice distortions, maintaining comparable deformation potentials for both electrons and holes (see Table~SV of SI~\cite{SI}). The combination of a softer lattice (low $\rho v_s^2$) and strongly coupled electronic states (large $D$) therefore creates a highly efficient acoustic scattering environment in the monolayer.

In bulk, the ADP channel reflects a balance between deformation coupling strength and the available scattering phase space. The larger deformation potential for holes ($D \approx 3.66$ eV) dominates over the higher electronic density of states for electrons (Sec.~IIIC(4)), yielding a hole relaxation time of $\sim200$ fs, nearly $50\%$ longer than for electrons. In contrast, monolayer transport exhibits a clear carrier-type asymmetry. For holes, the strong lattice coupling ($D \approx 3.54$ eV) dominates despite the relatively light effective mass ($m_h^* \approx 0.28m_e$), leading to a dramatic reduction of the acoustic lifetime to $\tau_{ADP}\approx14$ fs at $n=10^{18}$ cm$^{-3}$ (Fig.~\ref{fig:Tau_p}d). 
Electrons, however, experience a different balance. Their extremely small deformation potential ($D \approx 0.22$ eV) suppresses the scattering matrix element, although the heavier effective mass ($m_e^* \approx 0.34m_e$) increases the available phase space for scattering. As a result, the electron acoustic lifetime stabilizes at $\tau_{ADP}\approx22$ fs (Fig.~\ref{fig:Tau_n}d), about $60\%$ larger than the hole lifetime. This quantitative difference demonstrates that while a heavier effective mass enhances scattering probability, the small deformation potential remains the decisive factor preserving electron conductivity in the monolayer.

    The transport in bulk NaCrTe$_2$ is dominated by polar optical phonon (POP) scattering. The symmetric charge distribution on the Te sites ($+0.89e$, see Sec.~IIIC(2)) maintains moderate lattice ionicity, resulting in a strong Fröhlich coupling ($\alpha \approx 9.34 \times 10^{-3}$). Consequently, POP scattering sets the limiting carrier lifetime ($\tau_{\text{POP}} \approx 36$ and $\approx 40$ fs, Fig.~(\ref{fig:Tau_p},\ref{fig:Tau_n})b), while the acoustic deformation potential channel remains secondary ($\tau_{\text{ADP}} \approx 96$ fs and $\approx 195$ fs) for electrons and holes at room temperature. 
On the contrary, in the monolayer case, this hierarchy is reversed due to strong dielectric screening. Structural distortion suppresses the polar coupling: Bader analysis shows that the surface Te$_2$ site loses significant charge ($+0.51e$), reducing the in-plane Born effective charge ($Z^*_{\text{Cr}}$) by nearly 40\% (see Table SIII in SI\cite{SI}). Combined with confinement, this increases the static dielectric constant ($\varepsilon_0: 19.7 \rightarrow 44.5$). According to Eq.~\ref{eq:POP}, this enhanced screening strongly weakens long-range Coulomb interactions, causing the Fröhlich coupling to collapse by nearly an order of magnitude ($\alpha \approx 1.11 \times 10^{-3}$). As a result, the balance between ADP and POP mechanisms shifts, producing a fundamental ``scattering inversion'' upon dimensional reduction.

	\begin{figure*}
		\centering
		\includegraphics[width=1.0\linewidth]{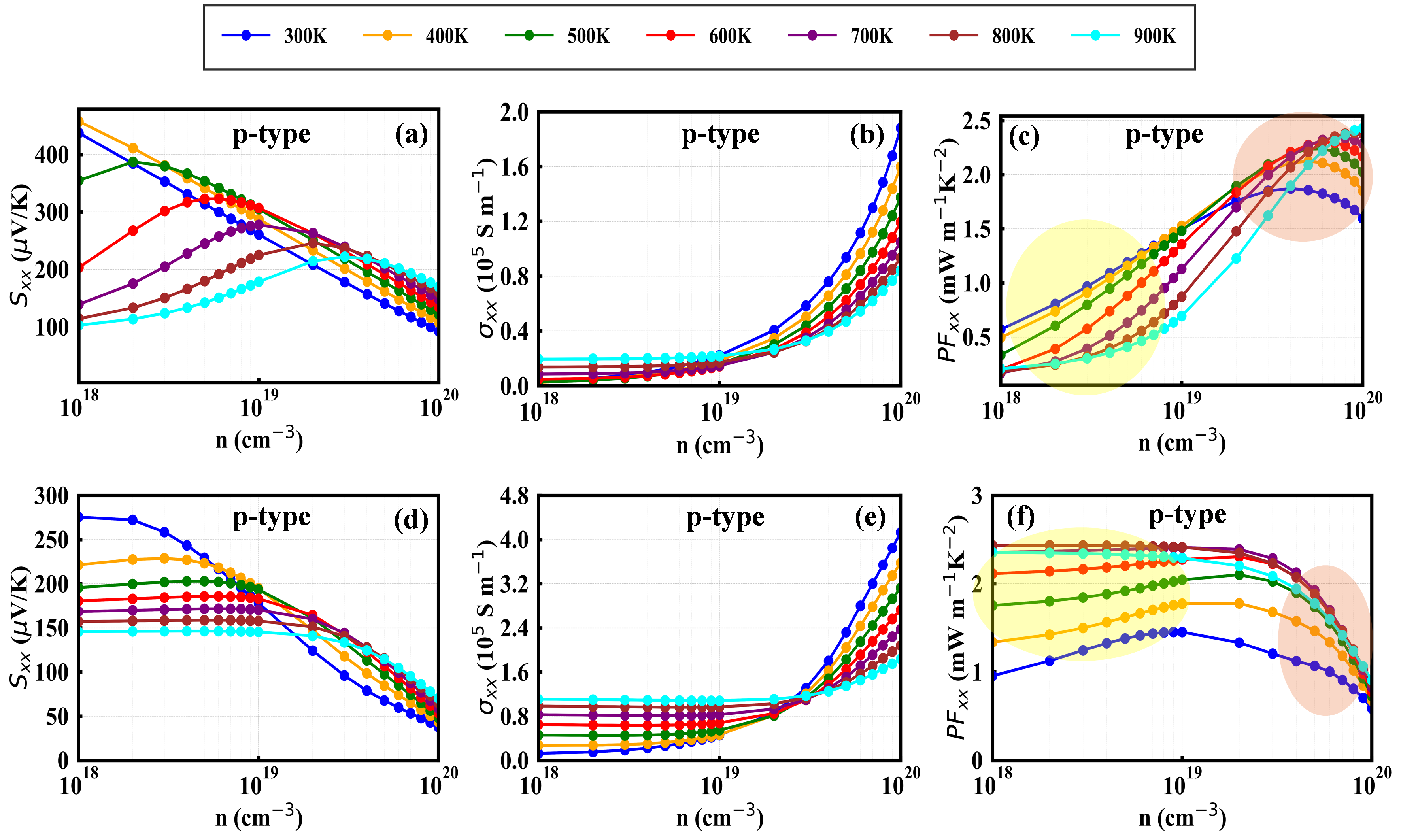}
		\caption{Temperature and carrier concentration dependence of Seebeck coefficient (S), electrical  conductivity ($\sigma$) and power-factor  (PF=S$^2\sigma$) of (a,b,c) bulk and (d,e,f) monolayer p-type NaCrTe$_2$. }
		\label{fig:Transport_properties_p}
	\end{figure*}
	
	\begin{figure*}
		\centering
		\includegraphics[width=1.0\linewidth]{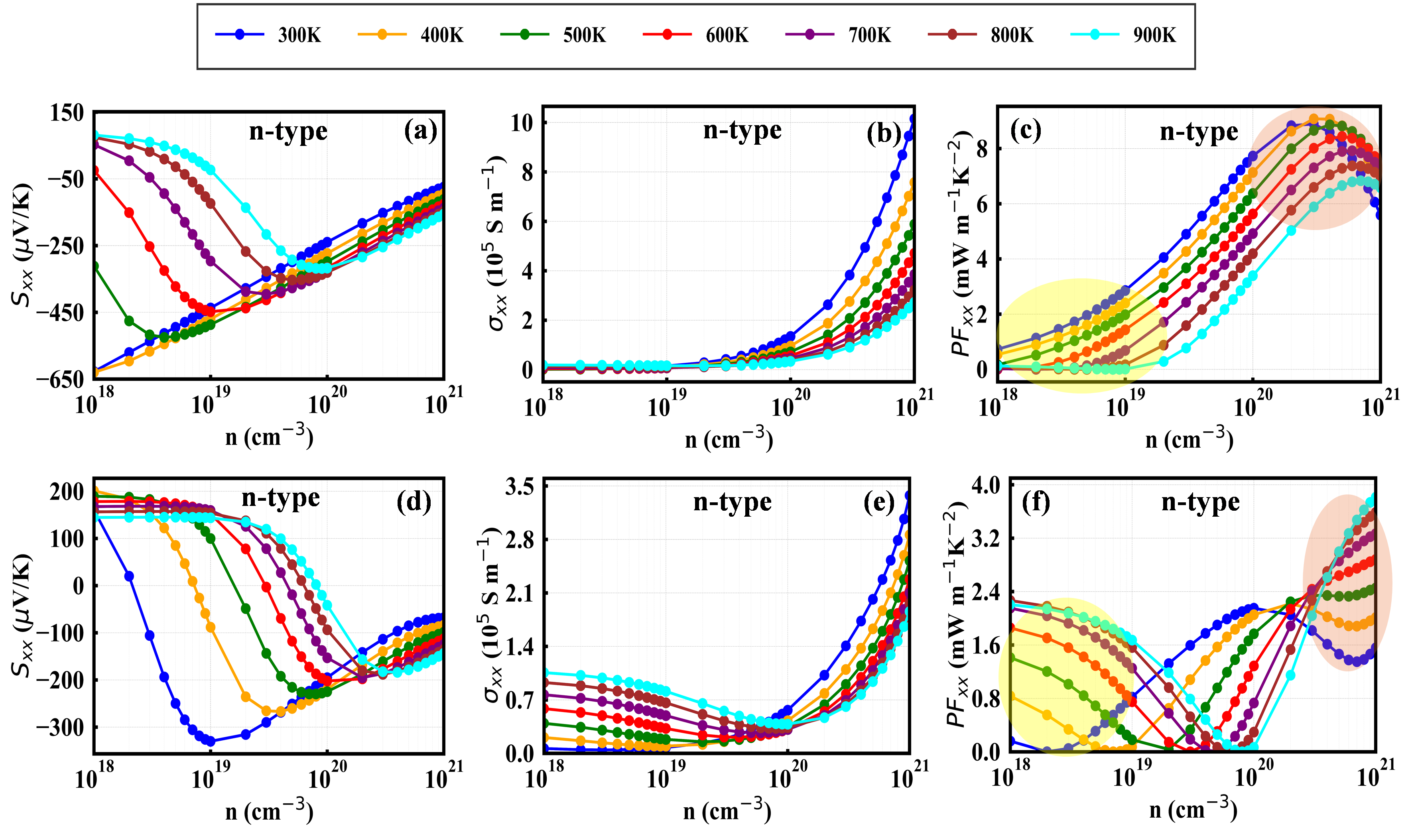}
		\caption{Temperature and carrier concentration dependence of Seebeck coefficient (S), electrical  conductivity ($\sigma$) and power-factor  (S$^2\sigma$) of (a,b,c) bulk and (d,e,f) monolayer n-type NaCrTe$_2$. }
		\label{fig:Transport_properties_n}
	\end{figure*}

The macroscopic transport coefficients ($S$, $\sigma$ and $S^2\sigma$) for bulk and monolayer systems are summarized in Fig.~\ref{fig:Transport_properties_p} and \ref{fig:Transport_properties_n}, revealing two distinct regimes depending on carrier concentration. At low concentrations, both systems exhibit pronounced bipolar effects. As evident from Fig.~\ref{fig:Transport_properties_p}(a) and \ref{fig:Transport_properties_n}(a), the Seebeck coefficient decreases with increasing temperature due to thermally excited minority carriers, particularly in the monolayer because of its narrow band gap. Despite its higher density of states, the monolayer shows a smaller Seebeck coefficient but benefits from enhanced electrical conductivity, yielding improved power factors at low concentration. For example, p-type monolayers reach a peak power factor of $\sim 2.5$ ${mW\,m}^{-1}{K}^{-2}$ at 900 K compared to $\sim 0.5$ ${mW\,m}^{-1}{K}^{-2}$ at 300 K (Fig.~\ref{fig:Transport_properties_p}(c,f)). A similar trend appears for n-type transport, where the monolayer achieves $\sim 2.3$ ${mW\,m}^{-1}{K}^{-2}$ at 900 K, exceeding the bulk value of $\sim 1.0$ ${mW\,m}^{-1}{K}^{-2}$ at 300 K.
At high carrier concentrations, the bipolar effect becomes negligible and the bulk system performs better due to its stronger Seebeck resilience. The bulk p-type system reaches $\sim 2.5$ ${mW\,m}^{-1}{K}^{-2}$ at 900 K, while n-type bulk exhibits an exceptional value of $\sim 9$ ${mW\,m}^{-1}{K}^{-2}$ near 400 K.

The n-type behavior reveals a contrasting physical picture. The bulk n-type system emerges as the high-performance candidate, reaching a peak power factor of $\approx 9.0$ ${mW\,m}^{-1}{K}^{-2}$ at 300 K (Fig.~\ref{fig:Transport_properties_n}(b)). This originates from the multi-valley nature of the bulk conduction band (see Sec.~III(C)), where high valley degeneracy simultaneously enhances the Seebeck coefficient ($\approx -80~\mu\text{V K}^{-1}$ at peak PF) and electrical conductivity. In contrast, the n-type monolayer shows lower conductivity than its p-type counterpart. Figure SX of Supporting Information\cite{SI} shows the Mobility variation for both bulk and monolayer NaCrTe$_2$. It indicates that while p-type monolayers maintain relatively high mobility despite shorter relaxation times, n-type monolayers exhibit reduced mobility due to stronger scattering.
This difference arises from the renormalization of the conductivity effective mass ($m^*_{\sigma}$), which is inversely proportional to mobility. The higher p-type mobility is associated with a lower conductivity effective mass. As shown in Table SVI, bands near the Fermi level exhibit larger velocities in the monolayer than in the bulk, indicating that p-type carriers in the 2D case experience reduced inertial resistance due to the smaller conductivity effective mass.

\textcolor{black}{The scattering-mechanism inversion, together with the confinement-induced band-gap reduction to 0.15~eV in the monolayer, suggests that NaCrTe$_2$ can serve as a versatile platform for multi-functional two-dimensional devices. In particular, the strong sensitivity of the Cr--Te $p$--$d$ hybridisation to structural reconstruction indicates that modest mechanical strains could efficiently tune both the band edges and the dominant scattering channel in the ADP-limited monolayer regime. This coupling between bonding, gap size, and carrier lifetimes points toward possible strain-engineered mid-infrared (MIR) optoelectronics\cite{MIR_photo, 2D_opto}, since the 0.15~eV gap lies in the relevant MIR window ($\sim 8~\mu$m), even though a full optoelectronic assessment is beyond the scope of the present study. Moreover, the evolution from an A-type AFM bulk to an FM monolayer, combined with the modest power-factor at low electron and hole concentrations, highlights NaCrTe$_2$ as a representative system for exploring two-dimensional spin-caloritronic physics. Additionally, confinement-driven changes in the gap and scattering hierarchy could also be exploited to control spin-polarized thermocurrents\cite{strain_1,strain_2}.
}

\section{Conclusions}
In this work, we present a comprehensive first-principles investigation of the impact of dimensionality reduction on the electronic, magnetic, and transport properties of the antiferromagnetic semiconductor $\text{NaCrTe}_2$. By systematically tracing the transition from bulk to the monolayer limit, we identify a coupled magnetostructural reconstruction that fundamentally alters the carrier dynamics. Quantum confinement breaks the inversion symmetry of the $\text{CrTe}_6$ octahedra, producing a disproportionation of Cr--Te bond lengths and strengthening the covalent character of bonding, as confirmed by Integrated Crystal Orbital Hamilton Population (ICOHP) and Bader charge analyses. This bonding evolution accompanies a magnetic transition from A-type antiferromagnetism in the bulk to ferromagnetism in the monolayer, with important consequences for electronic transport.

The structural reconstruction renormalizes the electronic structure, reducing the band gap from 0.44~eV in the bulk to 0.15~eV in the monolayer. This narrowing enhances bipolar transport at elevated temperatures, where thermally activated minority carriers partially suppress the Seebeck coefficient. Simultaneously, the reduction in bond ionicity substantially increases the static dielectric constant, with static dielectric constant ($\varepsilon_0$) rising from 19.7 to 44.5, accompanied by a strong suppression of the Born effective charges. In addition, the removal of interlayer coupling softens the lattice, lowering the longitudinal acoustic phonon velocity by approximately 30\%.
By combining these electronic and mechanical effects, we demonstrate that dimensionality reduction induces a complete inversion of the dominant carrier scattering mechanism. In the bulk, transport is governed by polar optical phonon (POP) scattering due to moderate lattice ionicity. In contrast, the enhanced dielectric screening in the monolayer strongly suppresses the Fr\"ohlich interaction, increasing the POP lifetime by nearly an order of magnitude. However, the accompanying lattice softening and increased density of states amplify acoustic deformation potential (ADP) scattering, which ultimately becomes the dominant transport bottleneck in the two-dimensional limit.

Our results reveal that while dielectric engineering through confinement effectively mitigates polar scattering, the concurrent mechanical softening imposes a fundamental constraint on carrier mobility. This resulting ``scattering inversion'' highlights the importance of scattering phase-space engineering in optimizing thermoelectric performance in magnetic van der Waals monolayers. Future strategies for improving 2D transport should therefore extend beyond dielectric screening and focus on band-structure modifications that preserve lattice stiffness or reduce conductivity effective masses to counteract the intrinsic limitations imposed by acoustic phonon scattering in reduced dimensions.
        
\section{Acknowledgements}
 H.S. acknowledges the financial support from the Indian Institute of Technology, Bombay, in the form of a teaching assistantship. H.S. also acknowledge the support of HPC facility PARAM Rudra under the National Supercomputing Mission.

		\bibliographystyle{unsrtnat}
		\bibliography{reference.bib}

	\end{document}